\documentclass[12pt]{article}
\usepackage{amsmath}
\usepackage{amssymb}
\usepackage{epsfig}
\usepackage{euscript}
\usepackage{fancybox}
\usepackage{color}

\def\mathswitchr#1{\relax\ifmmode{\mathrm{#1}}\else$\mathrm{#1}$\fi}

\newcommand {\pslash}{\hbox{$\not\hbox{\kern-2.3pt $p$}$}}

\usepackage{cite}

\usepackage{epic}

\def\alf1{ {\alpha\over\pi} }

\begin{document}
\begin{titlepage}
\begin{flushright}
 {\bf UTHEP-03-0301 }\\
{\bf Mar., 2003}\\
\end{flushright}
 
\begin{center}
{\Large Are Massive Elementary Particles Black Holes?$^{\dagger}$
}
\end{center}

\vspace{2mm}
\begin{center}
{\bf   B.F.L. Ward}\\
\vspace{2mm}
{\em Department of Physics and Astronomy,\\
  The University of Tennessee, Knoxville, Tennessee 37996-1200, USA}\\
\end{center}


\vspace{5mm}
\begin{center}
{\bf   Abstract}
\end{center}
We use exact results in a new approach to quantum gravity to study the effect
of quantum loop corrections on the behavior of the 
metric of space-time near the Schwarzschild radius of
a massive point particle in the Standard Model. We show that the classical
conclusion that such a system is a black hole is obviated.
Phenomenological implications are discussed.
\vspace{10mm}
\vspace{10mm}
\renewcommand{\baselinestretch}{0.1}
\footnoterule
\noindent
{\footnotesize
\begin{itemize}
\item[${\dagger}$]
Work partly supported 
by the US Department of Energy Contract  DE-FG05-91ER40627
and by NATO Grant PST.CLG.977751.
\end{itemize}
}

\end{titlepage}

\def\Kmax{K_{\rm max}}\def\ieps{{i\epsilon}}\def\rQCD{{\rm QCD}}
\renewcommand{\theequation}{\arabic{equation}}
\font\fortssbx=cmssbx10 scaled \magstep2
\renewcommand\thepage{}
\parskip.1truein\parindent=20pt\pagenumbering{arabic}\par
The classical theory of general relativity, as formulated by 
Albert Einstein, has had many successes~\cite{mtw,sw1}.
It has however so far evaded a complete, direct application 
of quantum mechanics, as all of the accepted treatments
of the complete quantum loop corrections to Einstein's theory involve
recourse to theoretical paradigms~\cite{gsw,jp} that are beyond 
the known phenomenology of the Standard Model ( SM )~\cite{sm}.
In Ref.~\cite{bw1}, we have introduced a new approach to
quantum gravity in which the apparently bad UV behavior of the 
theory is tamed by dynamical effects -- resummation of large 
higher order radiative corrections. This new approach,
which does not rely on phenomenologically unfounded theoretical
paradigms, allows us to compute finite quantum loop
effects and thus to analyze truly complete quantum effects in
Einstein's theory. In this paper, we present such an analysis.
\par

Among the most interesting of the outstanding questions, 
and there are many, posed by
Einstein's theory
to the quantum theory of point particle fields is the fate 
of massive point particles that are so crucial to the success of
the SM. In Einstein's theory, 
a point particle of non-zero rest mass $m$
has a non-zero Schwarzschild radius $r_S= 2(m/M_{Pl})(1/M_{Pl})$,
where $M_{Pl}$,~$1.22\times 10^{19}$ GeV, is the Planck mass,
so that such a particle should be a black hole~\cite{mtw} 
in the classical solutions
of Einstein's theory, unable to communicate ``freely'' with the world outside
of its Schwarzschild radius, except for some thermal effects
first pointed-out by Hawking~\cite{hawk}. 
Surely, this will not
do for the SM phenomenology, where it seems these point particles
are communicating freely their entire selves in their interactions with
each other. Can our new quantum theory of gravity reconcile
this apparent contradiction? It this question that we address in 
what follows.
\par

We start our analysis by setting up our new approach
to quantum gravity. As we explain in Ref.~\cite{bw1},
we follow the idea of Feynman~\cite{f1,f2} and treat
Einstein's theory as a point particle field theory
in which the metric of space-time undergoes quantum fluctuations
just like any other point particle does. On this view,
the Lagrangian density of the observable world is
\begin{equation}
{\cal L}(x) = -\frac{1}{2\kappa^2}\sqrt{-g} R
            + \sqrt{-g} L^{\cal G}_{SM}(x)
\label{lgwrld}
\end{equation}
where $R$ is the curvature scalar, $-g$ is the
negative of the determinant of the metric of space-time
$g_{\mu\nu}$, $\kappa=\sqrt{8\pi G_N}\equiv 
\sqrt{8\pi/M_{Pl}^2}$, where $G_N$ is Newton's constant,
and the SM Lagrangian density, which is well-known
( see for example, Ref.~\cite{sm,barpass} ) when invariance 
under local Poincare symmetry is not required,
is here represented by $L^{\cal G}_{SM}(x)$ which is readily obtained
from the familiar SM Lagrangian density as follows:
since $\partial_\mu\phi(x)$ is already generally
covariant for any scalar field $\phi$ and since the only derivatives of the
vector fields in the SM Lagrangian density occur in their
curls, $\partial_\mu A^J_\nu(x)-\partial_\nu A^J_\mu(x)$, which are
also already generally covariant, we only need
to give a rule for making the fermionic terms in 
usual SM Lagrangian density generally covariant. For this,
we introduce a differentiable structure with $\{\xi^a(x)\}$ as
locally inertial coordinates and an attendant
vierbein field $e^a_\mu\equiv\partial\xi^a/\partial x^\mu$ 
with indices that carry 
the vector representation for the flat locally inertial space, $a$, and for the
manifold of space-time, $\mu$, with the identification of the space-time
base manifold metric as
$g_{\mu\nu}=e^a_\mu e_{a\nu}$ where the flat locally inertial 
space indices are to be
raised and lowered with Minkowski's metric $\eta_{ab}$ as usual. 
Associating the usual Dirac gamma
matrices $\{\gamma_a\}$ with the flat locally inertial space at x, we define
base manifold Dirac gamma matrices by $\Gamma_\mu(x)=e^a_\mu(x)\gamma_a$.
Then the spin connection, $\omega_{\mu b}^a=-\frac{1}{2}e^{a\nu}\left(
\partial_\mu e^b_\nu-\partial_\nu e^b_\mu\right)+\frac{1}{2}e^{b\nu}\left(
\partial_\mu e^a_\nu-\partial_\nu e^b_\mu\right)
+\frac{1}{2}e^{a\rho}e^{b\sigma}\left(\partial_\rho e_{c\sigma}-\partial_\sigma e_{c\rho}\right)e^c_\mu$ when there is no torsion, allows us to 
identify the generally covariant
Dirac operator for the SM fields by the substitution
${i\not{\partial}} \rightarrow i\Gamma(x)^\mu\left(\partial_\mu +\frac{1}{2}{\omega_{\mu b}}^a{\Sigma^b}_a\right)$, where we have ${\Sigma^b}_a=\frac{1}{4}\left[\gamma^b,\gamma_a\right]$
everywhere in the SM Lagrangian density. This will generate $L^{\cal G}_{SM}(x)$ from the usual SM Lagrangian density $L_{SM}(x)$ as it is
given in Refs.~\cite{sm,barpass}, for example.\par

It is well-known that there are many massive 
point particles in (\ref{lgwrld}).
According to classical general relativity, they should all be black holes,
as we noted above. Are they black holes in our new approach to quantum gravity?
To study this question, we continue to follow Feynman in Ref.~\cite{f1,f2}
and treat spin as an inessential complication~\cite{mlg}, 
as the question of whether
a point particle with mass is or is not a black hole should not depend
too severely on whether or not it is spinning. We can come back to a 
spin-dependent analysis elsewhere~\cite{elsewh}.\par 

Thus, we replace $L^{\cal G}_{SM}(x)$ in (\ref{lgwrld})
with the simplest case for our question, that of a free scalar field
, a free physical Higgs field, $\varphi(x)$, with a rest mass believed~\cite{lewwg} to be less than $400$ GeV and known to be greater than $114.4$ GeV with a
95\% CL. We are then led to consider the representative model
\begin{equation}
\begin{split}
{\cal L}(x) &= -\frac{1}{2\kappa^2} R \sqrt{-g}
            + \frac{1}{2}\left(g^{\mu\nu}\partial_\mu\varphi\partial_\nu\varphi - m_o^2\varphi^2\right)\sqrt{-g}\\
            &= \quad \frac{1}{2}\left\{ h^{\mu\nu,\lambda}\bar h_{\mu\nu,\lambda} - 2\eta^{\mu\mu'}\eta^{\lambda\lambda'}\bar{h}_{\mu_\lambda,\lambda'}\eta^{\sigma\sigma'}\bar{h}_{\mu'\sigma,\sigma'} \right\}\\
            & \qquad + \frac{1}{2}\left\{\varphi_{,\mu}\varphi^{,\mu}-m_o^2\varphi^2 \right\} -\kappa {h}^{\mu\nu}\left[\overline{\varphi_{,\mu}\varphi_{,\nu}}+\frac{1}{2}m_o^2\varphi^2\eta_{\mu\nu}\right]\\
            & \quad - \kappa^2 \left[ \frac{1}{2}h_{\lambda\rho}\bar{h}^{\rho\lambda}\left( \varphi_{,\mu}\varphi^{,\mu} - m_o^2\varphi^2 \right) - 2\eta_{\rho\rho'}h^{\mu\rho}\bar{h}^{\rho'\nu}\varphi_{,\mu}\varphi_{,\nu}\right] + \cdots \\
\end{split}
\label{eq1}
\end{equation}
Here, 
$\varphi(x)_{,\mu}\equiv \partial_\mu\varphi(x)$,
and $g_{\mu\nu}(x)=\eta_{\mu\nu}+2\kappa h_{\mu\nu}(x)$ 
where we follow Feynman and expand about Minkowski space
so that $\eta_{\mu\nu}=diag\{1,-1,-1,-1\}$. 
Following Feynman, we have introduced the notation
$\bar y_{\mu\nu}\equiv \frac{1}{2}\left(y_{\mu\nu}+y_{\nu\mu}-\eta_{\mu\nu}{y_\rho}^\rho\right)$ for any tensor $y_{\mu\nu}$\footnote{Our conventions for raising and lowering indices in the 
second line of (\ref{eq1}) are the same as those
in Ref.~\cite{f2}.}. 
Thus, $m_o$ is the bare mass of our free Higgs field and we set the small
tentatively observed~\cite{cosm1} value of the cosmological constant
to zero so that our quantum graviton has zero rest mass.
The Feynman rules for (\ref{eq1}) have been essentially worked out by 
Feynman~\cite{f1,f2}, including the rule for the famous
Feynman-Faddeev-Popov~\cite{f1,ffp1} ghost contribution that must be added to
it to achieve a unitary theory with the fixing of the gauge
( we use the gauge of Feynman in Ref.~\cite{f1}, 
$\partial^\mu \bar h_{\nu\mu}=0$ ), 
so we do not repeat this 
material here. We turn instead directly to the issue 
of the effect of quantum loop corrections
on the black hole character of our massive Higgs field.
\par

\begin{figure}
\begin{center}
\epsfig{file=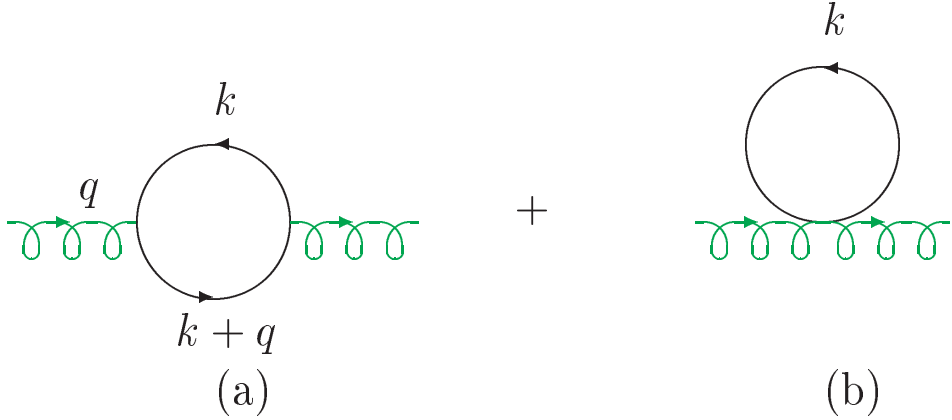,width=140mm}
\end{center}
\caption{\baselineskip=7mm     The scalar one-loop contribution to the
graviton propagator. $q$ is the 4-momentum of the graviton.}
\label{fig1}
\end{figure}

To initiate our approach, let us study the possible one-loop corrections to
Newton's law that would follow from the matter in (\ref{eq1}).
We will show that these corrections directly impact our black hole issue.
It is sufficient to calculate the effects of the diagrams
in Fig.~\ref{fig1} on the graviton 
propagator to see the first quantum loop effect.
\par

In Ref.~\cite{bw1}, we have shown that, while
the naive power counting of
the graphs gives their degree of divergence as +4,
YFS~\cite{yfs1} resummation of the soft graviton effects in the propagators
in Fig.~\ref{fig1} renders the graphs ultra-violet (UV) finite. 
Indeed, for example, for
Fig. 1a, we get without YFS resummation the result 
\begin{equation}
i\Sigma(q)^{1a}_{\bar\mu\bar\nu;\mu\nu}=\kappa^2\frac{\int d^4k}{2(2\pi)^4}
\frac{\left(k'_{\bar\mu}k_{\bar\nu}+k'_{\bar\nu}k_{\bar\mu}\right)
\left(k'_{\mu}k_{\nu}+k'_{\nu}k_{\mu}\right)}
{\left({k'}^2-m^2+i\epsilon\right)\left(k^2-m^2+i\epsilon\right)}
\label{eq2}
\end{equation},
where we set $k'=k+q$ and we take for definiteness only
fully transverse, traceless polarization states of the graviton to
be act on $\Sigma$ so that we have dropped the traces from its
vertices. Clearly, (\ref{eq2})
has degree of divergence +4. When we take into 
account the resummation as calculated in Ref.~\cite{bw1},
the free scalar propagators are improved to their YFS-resummed values,
\begin{equation}
i\Delta'_F(k)|_{YFS-resummed} =  \frac{ie^{B''_g(k)}}{(k^2-m^2+i\epsilon)}
\label{resum}
\end{equation},
where the virtual graviton function $B''_g(k)$ is, for Euclidean momenta,
\begin{equation}
B''_g(k) = \frac{\kappa^2|k^2|}{8\pi^2}\ln\left(\frac{m^2}{m^2+|k^2|}\right),
\label{deep}
\end{equation}  
so that we get instead of (\ref{eq2}) the result
( here, $k\rightarrow (ik^0,\vec k)$ by Wick rotation )
\begin{equation}
i\Sigma(q)^{1a}_{\bar\mu\bar\nu;\mu\nu}=i\kappa^2\frac{\int d^4k}{2(2\pi)^4}
\frac{\left(k'_{\bar\mu}k_{\bar\nu}+k'_{\bar\nu}k_{\bar\mu}\right)e^{\frac{\kappa^2|{k'}^2|}{8\pi^2}\ln\left(\frac{m^2}{m^2+|{k'}^2|}\right)}
\left(k'_{\mu}k_{\nu}+k'_{\nu}k_{\mu}\right)e^{\frac{\kappa^2|k^2|}{8\pi^2}\ln\left(\frac{m^2}{m^2+|k^2|}\right)}}
{\left({k'}^2-m^2+i\epsilon\right)\left(k^2-m^2+i\epsilon\right)}.
\label{eq2p}
\end{equation}
Evidently, this integral converges; so does that for Fig.1b when
we use the improved resummed propagators. This means that we
have a rigorous quantum loop correction to Newton's law
from Fig.1 which is finite and well defined.\par

To see how this result impacts the black hole character of our
massive point particle, we continue to work in the transverse, traceless
space for the graviton self-energy $\Sigma$\footnote{ As all physical polarization states are propagated with the same Feynman denominator, any physical
subspace can be used to determine this denominator. } and  
we get, to leading order, that the graviton propagator denominator
becomes
\begin{equation}
q^2 +\frac{1}{2}q^4\Sigma^{T(2)}+i\epsilon
\label{prop}
\end{equation}
where the transverse, traceless self-energy function $\Sigma^T(q^2)$ follows from eq.(\ref{eq2p}) for Fig. 1a and its analog for Fig. 1b
by the standard methods. 
For the coefficient of $q^4$ in $\Sigma^T(q^2)$ for $|q^2|>>m^2$
we have the result
\begin{equation} 
-\frac{1}{2}\Sigma^{T(2)} \cong \frac{c_2}{360\pi M_{Pl}^2}
\label{sigma}
\end{equation}
for
\begin{equation} 
c_2 = \int^{\infty}_{0}dx x^3(1+x)^{-4-\lambda_c x}\cong 72.1 
\label{int1}
\end{equation}
where $\lambda_c=\frac{2m^2}{\pi M_{Pl}^2}$.
When we Fourier transform the
inverse of (\ref{prop}) we find the potential
\begin{equation}
\Phi_{Newton}(r)= -\frac{G_N M_1M_2}{r}(1-e^{-ar})
\end{equation}
where $a=1/\sqrt{-\frac{1}{2}\Sigma^{T(2)}}\simeq 3.96 M_{Pl}$
in an obvious notation, where for definiteness, we set $m\cong 120$GeV.\par

At this point, let us note that the integral in (\ref{int1}) can be represented
for our purposes by the analytic expression~\cite{elsewh}
\begin{equation}
c_2 \cong \ln\frac{1}{\lambda_c}-\ln\ln\frac{1}{\lambda_c}-\frac{\ln\ln\frac{1}{\lambda_c}}{\ln\frac{1}{\lambda_c}-\ln\ln\frac{1}{\lambda_c}}-\frac{11}{6}
\label{anal1}
\end{equation}
and we used this result to check the numerical result given in (\ref{int1}).
It is clear that, without resummation, we would have
$\lambda_c=0$ and our result in (\ref{int1})
would be infinite and, since this is the coefficient of $q^4$ in 
the inverse propagator,
no renormalization of the field and of the mass could be used to remove
such an infinity. In our new approach to quantum gravity, this 
infinity is absent.\par

We stress that our result in (\ref{sigma}) is gauge invariant, as our 
approach involves the exact re-arrangement of the Feynman series
as we explain in Ref.~\cite{bw1} and the original series is gauge
invariant. Indeed, one can cross check this result by comparing with
the pioneering work in Ref.~\cite{thvelt1}, where the complete
result of the one-loop divergences of our scalar field coupled
to Einstein's gravity have been computed. This is made possible by the
following observation. As we just observed, the result which we
have obtained would be UV divergent without our resummation. Thus,
the dominant terms which we are isolating in this paper
are precisely those that
are given in Ref.~\cite{thvelt1}, where we need to make the correspondence
between the poles in $n$, the dimension of space-time,
at $n=4$ calculated in Ref.~\cite{thvelt1} and the leading 
log $\ln\frac{1}{\lambda_c}$. This we do by setting 
the result $c_2$ equal to its value when $\lambda_c = 0$ in $n$ dimensions
and allowing $n \rightarrow 4$. In this way we find that
\begin{equation}
1/(2-n/2) \leftrightarrow c_2 .
\label{crsp1}
\end{equation}
This means that, if we look at the limit $q^2\rightarrow 0$,
we get the result that the coefficient of $q^4$ 
in  (\ref{prop}) is $3/(2-n/2)$ times the coefficient 
of $c_2$ on the right-hand side of (\ref{sigma}), and this is in
complete agreement with the result that is implied by
eq.(3.40) in  Ref.~\cite{thvelt1}, for example. Of course, the results in
Ref.~\cite{thvelt1} are also gauge invariant.\par

In the SM, there are now believed to be three massive neutrinos~\cite{neut},
with masses that we estimate at $\sim 3$ eV, and the remaining members
of the known three generations of Dirac fermions $\{e,\mu,\tau,u,d,s,c,b,t\}$, 
with masses given by ~\cite{pdg2002},
$m_e\cong 0.51$ MeV, $m_\mu \cong 0.106$ GeV, $m_\tau \cong 1.78$ GeV,
$m_u \cong 5.1$ MeV, $m_d \cong 8.9$ MeV, $m_s \cong 0.17$ GeV,
$m_c \cong 1.3$ GeV, $m_b \cong 4.5$ GeV and $m_t \cong 174$ GeV,
as well as the massive vector bosons $W^{\pm},~Z$, with masses 
$M_W\cong 80.4$ GeV,~$M_Z\cong 91.19$ GeV. Using the general spin independence
of the graviton coupling to matter at {\it relatively low} momentum transfers,
we see that we can take the effects of these degrees of freedom into account
approximately by counting each Dirac fermion as 4 degrees of freedom,
each massive vector boson as 3 degrees of freedom and remembering that
each quark has three colors. Using the result (\ref{anal1}) for each
of the massive degrees of freedom in the SM, we see that
the effective value of $c_2$ in the SM is approximately
\begin{equation}
c_{2,eff} \cong 9.26\times 10^3
\label{ceff} 
\end{equation}
so that the effective value of $a$ in the SM is 
\begin{equation}
a_{eff} \cong 0.349 M_{Pl} .
\label{aeff} 
\end{equation}
To make direct contact with black hole physics, note that,
for $r\rightarrow r_S$, $a_{eff}r \ll 1$ so 
that $|2\Phi_{Newton}(r)|_{M_1=m}/M_2|\ll 1$.
This means that in the respective solution for our metric of space-time,
$g_{00}\cong 1+2\Phi_{Newton}(r)|_{M_1=m}/M_2$ remains 
positive as we pass through the
Schwarzschild radius. Indeed, it can be shown that this 
positivity holds to $r=0$. Similarly, $g_{rr}$ remains negative
through $r_S$ down to $r=0$. To get these results,
note that in the relevant regime for r, the smallness of
the quantum corrected Newton potential means that we can use the
linearized Einstein equations for a small spherically symmetric
static source $\rho(r)$ which generates $\Phi_{Newton}(r)|_{M_1=m}/M_2$
via the standard Poisson's equation. 
The usual result~\cite{abs,mtw,sw1} for the
respective metric solution then gives 
$g_{00}\cong 1+2\Phi_{Newton}(r)|_{M_1=m}/M_2$ and
$g_{rr}\cong -1+2\Phi_{Newton}(r)|_{M_1=m}/M_2$ which remain
respectively time-like and space-like
to $r=0$.\par
It follows that the quantum corrections have obviated the classical
conclusion that a massive point particle is a black hole~\cite{mtw}.\par

We do not wish to suggest that the value of $a_{eff}$ given here is complete,
as there may be as yet unknown massive particles beyond those already
discovered. Including more particles in the computation of
$a_{eff}$ would make it smaller and hence would not change the
conclusions of our analysis. For example, in the Minimal 
Supersymmetric Standard
Model we expect approximately that 
$a_{eff}\rightarrow \frac{1}{\sqrt{2}}a_{eff}$.
In addition, we point-out that, using the correspondence
in (\ref{crsp1}) one can also use the results for the
complete one-loop corrections in Ref.~\cite{thvelt1}
to the theory treated here to see that the remaining 
interactions at one-loop order not discussed here 
(vertex corrections, pure gravity self-energy corrections, etc. )
also do not increase the value of $a_{eff}$~\cite{elswh}.
We can thus think of $a_{eff}$
as a parameter which is bounded from above by the estimates we give
above and which should be determined from cosmological and/or other
considerations. Further such implications will be taken up elsewhere.\par

\section*{Acknowledgements}

We thank Profs. S. Bethke and L. Stodolsky for the support and kind
hospitality of the MPI, Munich, while a part of this work was
completed. We thank Prof. C. Prescott for
the kind hospitality of SLAC Group A while this
work was in progress. We thank Prof. S. Jadach for useful discussions.


\newpage
\begin{thebibliography}{99}
\bibitem{mtw} C. Misner, K.S. Thorne and J.A. Wheeler,
{\it Gravitation},( Freeman, San Francisco, 1973 ).
\bibitem{sw1} S. Weinberg, {\it Gravitation and Cosmology: Principles and Applications of the General Theory of Relativity},( John Wiley, New York, 1972).
\bibitem{gsw} See, for example, M. Green, J. Schwarz and E. Witten,
{\em Superstring Theory, v. 1 and v.2}, ( Cambridge Univ. 
Press, Cambridge, 1987 ) and references therein.
\bibitem{jp}
See, for example, J. Polchinski, {\em String Theory, v. 1 and v. 2},
(Cambridge Univ. Press, Cambridge, 1998), and references therein.
\bibitem{sm}S.L. Glashow, Nucl. Phys. {\bf 22} (1961) 579; 
S. Weinberg, Phys. Rev. Lett. {\bf 19} (1967) 1264;
A. Salam, in {\em Elementary Particle Theory}, ed. N. Svartholm
(Almqvist and Wiksells, Stockholm, 1968), p. 367;
G.~'t Hooft and M.~Veltman, Nucl. Phys. {\bf B44},189 (1972)
and {\bf B50}, 318 (1972); 
G.~'t Hooft, {\it ibid.} {\bf B35}, 167 (1971); M.~Veltman, {\it ibid.} {\bf B7}, 637 (1968); 
D. J. Gross and F. Wilczek, 
Phys. Rev. Lett. {\bf 30} (1973) 1343;
H. David Politzer, {\it ibid.}{\bf 30} (1973) 1346; see also
, for example, F. Wilczek, in {\em Proc. 16th International Symposium on Lepton and 
Photon Interactions, Ithaca, 1993}, eds. P. Drell and D.L. Rubin 
(AIP, NY, 1994) p. 593, and references therein.
\bibitem{bw1} B.F.L. Ward, Mod. Phys. Lett. A{\bf 17} (2002) 2371.
\bibitem{hawk} S. Hawking, Nature ( London ) {\bf 248} (1974) 30;
Commun. Math. Phys. {\bf 43} ( 1975 ) 199.
\bibitem{f1} R. P. Feynman, Acta Phys. Pol. {\bf 24} (1963) 697.
\bibitem{f2} R. P. Feynman, {\em Feynman Lectures on Gravitation},
eds. F.B. Moringo and W.G. Wagner (Caltech, Pasadena, 1971).
\bibitem{barpass} D. Bardin and G. Passarino,{\it The Standard 
Model in the Making : Precision Study of the Electroweak Interactions },
( Oxford Univ. Press, London, 1999 ).
\bibitem{mlg} M.L. Goldberger, private communication, 1972.
\bibitem{elsewh} B.F.L. Ward, to appear.
\bibitem{lewwg}
D. Abbaneo {\it et al.}, hep-ex/0212036; see also, M. Gruenewald, 
hep-ex/0210003, in {\it Proc. ICHEP02}, in press, 2003.
\bibitem{cosm1}S. Perlmutter {\it et al.}, Astrophys. J. {\bf 517} (1999) 565;
and, references therein.
\bibitem{ffp1} L. D. Faddeev and V.N. Popov, ITF-67-036, 
NAL-THY-57 (translated from Russian by D. Gordon and B.W. Lee);
Phys. Lett. {\bf B25} (1967) 29.
\bibitem{yfs1}D.~R.~Yennie, S.~C.~Frautschi, and H.~Suura, Ann. Phys. {\bf 13} (1961) 379;\newline
see also K.~T.~Mahanthappa, {\sl Phys.~Rev.~\bf 126} (1962) 329, for a related analysis.
\bibitem{thvelt1}
G. 't Hooft and M. Veltman, Ann. Inst. Henri Poincare {\bf XX}, 69 (1974).
\bibitem{neut} See for example D. Wark, in {\it Proc. ICHEP02}, in press; and,
M. C. Gonzalez-Garcia, hep-ph/0211054, in {\it Proc. ICHEP02}, in press,
and references therein.
\bibitem{pdg2002}K. Hagiwara {\it et al.}, Phys. Rev. D{\bf 66} (2002) 010001;
see also H. Leutwyler and J. Gasser, Phys. Rept. {\bf 87} (1982) 77, and
references therein.
\bibitem{abs} R. Adler, M. Bazin and M. Schiffer, {\it Introduction to General
Relativity },( McGraw-Hill, New York, 1965 ).
\bibitem{elswh}
B.F.L. Ward, to appear.
\end{thebibliography}
\end{document}